\documentclass[journal]{IEEEtran}
\usepackage{cite}

\ifCLASSINFOpdf 
\usepackage[pdftex]{graphicx}
\else
\usepackage[dvips]{graphicx}
\fi

\usepackage[cmex10]{amsmath}
\usepackage{algorithmic}
\usepackage[ruled,norelsize]{algorithm2e}

\usepackage{array}

\usepackage{mdwmath}
\usepackage{mdwtab}
\usepackage{todonotes}

\usepackage{eqparbox}

\ifCLASSOPTIONcompsoc
\usepackage[caption=false,font=normalsize,labelfont=sf,textfont=sf]{subfig}
\else
\usepackage[caption=false,font=footnotesize]{subfig}
\fi

\usepackage{fixltx2e}
\usepackage{url}
\usepackage{graphicx}
\usepackage{amsfonts}
\usepackage{amssymb}
\usepackage{amsthm}
\usepackage{bm}
\usepackage[ruled]{algorithm2e}
\usepackage[utf8]{inputenc}
\usepackage{booktabs}
\usepackage{epstopdf}
\usepackage{cite}
\usepackage{flushend}

\newcounter{tempEquationCounter} 
\newcounter{thisEquationNumber}

\begin{document}

%
\title{Comments on ``CRB-RPL: A Receiver-based Routing Protocol for Communications in \\Cognitive Radio Enabled Smart Grid''}

\author{Adnan~Aijaz,~\IEEEmembership{Member,~IEEE}
\thanks{The author is with the Telecommunications Research Laboratory, Toshiba Research Europe Ltd., Bristol, BS1 4ND, UK. Contact e-mail: adnan.aijaz@toshiba-trel.com}}
\maketitle
\begin{abstract}
\boldmath
A recent paper by Yang \emph{et al.} \cite{R_smelly} proposed CRB-RPL as a new RPL-based routing protocol for communications in cognitive radio (CR) enabled smart grid. Essentially, CRB-RPL adopts the operation of CRB-MAC, which is a medium access control (MAC) protocol for cognitive machine-to-machine (M2M) communications, and depicts it as a network layer enhancement. CRB-RPL suffers from a number of technical flaws in terms of protocol operation and analytical aspects. The main objective of this paper is to highlight these technical flaws of CRB-RPL and provide corrections to analytical modeling. 

\end{abstract}



%
\IEEEpeerreviewmaketitle

\section{Introduction}
\IEEEPARstart{I}{n} a recent paper \cite{R_smelly}, Yang \emph{et al.} proposed CRB-RPL as a new RPL\footnote{Routing Protocol for Low Power and Lossy Networks (RPL) \cite{RPL}}-based routing protocol for communications in cognitive radio (CR) enabled smart grid. CRB-RPL draws inspiration from two recent protocols for cognitive machine-to-machine (M2M) communications \cite{R_CM2M}: (i) CRB-MAC \cite{R_CRB}, which is a medium access control (MAC) protocol for cognitive M2M applications, and (ii) CORPL \cite{R_CORPL}, which is a routing protocol for cognitive M2M applications. The key aspect of CRB-MAC is a \emph{receiver-based} forwarding mechanism at the MAC layer, which improves the reliability, particularly in lossy environments. On the other hand, CORPL provides novel enhancements to RPL for operation in CR environments. CORPL adopts an opportunistic forwarding approach at the network layer for meeting the utility requirements of the secondary network along with ensuring protection to the primary users. 

CRB-RPL adopts the receiver-based forwarding approach of CRB-MAC and depicts it as a network layer protocol. It should be noted that CRB-MAC is a cross-layer MAC protocol wherein the network layer, which is not necessarily limited to RPL, assigns a rank to each node, based on its virtual distance from the sink/gateway node. Such rank is used by a node to take part in the election process for forwarding the packet. CRB-RPL essentially achieves the same operation by  following the CRB-MAC approach with minor modifications. Therefore, the authors' claim of a \emph{new} (novel) protocol is not justified, due to significant overlap of CRB-RPL with CRB-MAC, which is evident in terms of its protocol operation and analytical aspects. CRB-RPL also aims to achieve CORPL-like functionality, in terms of providing protection to primary receivers; however, this creates some issues as discussed later. Moreover, CRB-RPL suffers from a number of technical flaws, in terms of protocol operation and analytical modeling. It is the aim of this paper  to highlight these flaws and provide corrections, wherever applicable.


\section{Comments on Protocol Operation }
The main comments on the protocol operation of CRB-RPL are given as follows.

\begin{enumerate}
\item 
The authors claim that CRB-RPL is an \emph{RPL-based} routing protocol. However, this is not true. The fundamental principle of RPL operation is to maintain network state information in the form of one or more \emph{directed acyclic graphs} (DAGs). Each node in the DAG is assigned a rank which is computed based on an objective function. It should be noted that rank computation is not specific to RPL. In RPL, the traffic generated by a node is forwarded to the next hop through its default parent. However, by adopting a receiver-based approach, multiple nodes in the neighborhood will \emph{independently} take part in the forwarding process. Hence, the DAG structure is not retained any more. This is unlike CORPL, which retains the DAG structure despite adopting an opportunistic forwarding approach, through prioritization of nodes in the forwarding set, including the default parent, by the sender node. 

\item
One of the key objectives of CRB-RPL is to ensure protection to primary receivers by selecting next hop nodes having minimum overlap with coverage of the primary transmitters. It should be noted that ensuring protection to primary receivers at the network layer is a key feature of CORPL as well. In CRB-RPL, this is achieved by introducing the \emph{cognitive transmission quality} (CTQ) metric, given by (6) in \cite{R_smelly}, which contains a factor accounting for the aforementioned coverage overlap. The CTQ is used in rank computation by a node based on (8) of \cite{R_smelly}. However, protection to primary receivers is not guaranteed by such rank computation alone. As mentioned in the previous comment, the DAG structure in CRB-RPL is not retained due to the receiver-based forwarding approach, and therefore, the sender node has no control over selecting the next hop. Moreover, CRB-RPL does not describe how a node with lower coverage with primary transmitters will have priority in forwarding the packet. Furthermore, the CTQ metric is affected by the link success probability between a sender and a receiver node. Therefore, the gain provided by a node in terms of lower coverage overlap can be compromised by  poor link quality. It should be noted CORPL guarantees primary receiver protection by prioritizing nodes in the forwarder set based on their coverage overlap with primary transmitters. The priority information is  embedded into the MAC layer header of the forwarded packet.

\end{enumerate}


\section{Comments on Analytical Aspects }
The main comments on analytical aspects of CRB-RPL are given as follows. 

\begin{enumerate}
\item 
The authors define hop energy efficiency (HEE) as a ratio of the hop distance between two nodes and the total energy consumption to forward to the packet, which is given by (11) of \cite{R_smelly}. The authors further calculate the hop distance as the rank difference between the sender and receiver nodes, which is given by (12) of \cite{R_smelly}. However, such method for computing hop distance is incorrect. Consider the scenario shown in authors' own example of Fig. 2. Based on (12) of \cite{R_smelly}, the hop distance of node 5 from node 1 comes out to be \(4.04 - 1 = 3.04  \); however, node 5 is actually 2 hops away from the gateway. Such inaccuracy is a direct consequence of determining hop distance based on rank of the nodes. It should be noted that the rank, in context of RPL, is generally computed based on link success probability, which is a highly dynamic parameter. Therefore, results for hop distance will vary significantly at different time instants. An appropriate approach to compute hop distance is to assign a unique hop ID to all the nodes which are equidistant (in terms of hops) from the gateway node and compute the hop distance based on difference in hop IDs of the nodes. Such hop ID can be easily assigned during the  association phase. 

\item 
The parameters \(\delta^2\) and \(\mathbb{P}_{acc}^k\) in (20) of \cite{R_smelly}, which computes the minimum required transmit power,  are undefined throughout the paper. For technical correctness, \(\delta^2\) must represent the noise power (frequently denoted by \(\sigma^2\) in literature). Moreover, \(\mathbb{P}_{acc}^k\) must be replaced by the potential bandwidth of the \(k^{th}\) channel, which is given by \(\mathbb{P}_{acc}^k \cdot B_k\), where \(B_k\) is the actual bandwidth of the \(k^{th}\) channel and \(\mathbb{P}_{acc}^k\) denotes the probability of switching transmission to the \(k^{th}\) channel, which is given by (7) of \cite{R_CRB}.

\item
The average number of retransmissions until success for CRB-RPL, denoted by \(\chi\), is given in (25) of \cite{R_smelly}. However, the equation is not given in closed-form, which can be obtained by (15) of \cite{R_CRB}. 

\item
In Section IV.C, the authors evaluate the \emph{coordination overhead}\footnote{The coordination overhead \cite{R_CORPL} is defined as the probability of a node in the forwarder set retransmitting a packet when any other node has already forwarded it to the next hop.} of CRB-RPL, given by (32) -- (33) of \cite{R_smelly}. Such computation of coordination overhead is inaccurate not only in terms of approach but also in terms of derived expressions. Intuitively, the coordination overhead should account for the erroneous (duplicate) forwarding of a packet at any hop along with the probability of receiving this packet at the gateway in a multi-hop manner. The single-hop coordination overhead, given by (32), only accounts for error in preamble transmission. Besides, (32) is not correct as the probability of transmission on a cognitive channel, \(P_{sw}^j\), is independent of the probability of error in preamble transmission. Moreover, an erroneous forwarding of packet will take place if a receiving node does not hear a preamble or a data frame on the medium. Therefore, in addition to the aforementioned inaccuracy, (32)  does not account for all possible factors involved in the protocol operation. Most importantly, the coordination overhead of the route (which is intuitively and technically wrong to compute), given by (33), does not account for the reception of duplicate packet at the gateway. Finally, the coordination overhead, as computed by the authors, decreases as the number of receiver nodes in the forwarder set increases. However, this is not intuitive, as it violates the principles of opportunistic forwarding. The correct expression for coordination overhead is derived in the next section.   


\end{enumerate}

\section{Calculation of Coordination Overhead}
For the sake of calculating the coordination overhead, we adopt the same notation as employed in \cite{R_CRB} and \cite{R_smelly}. Let, $Y_a=1/P_{ab}$ denote the cost of forwarding a message from node $a$ to node $b$, where $P_{ab}=1-\mathcal{P}_f$ is the probability of successful transmission, such that  $\mathcal{P}_f$ is the probability of failure for a transmission between nodes $a$ and $b$. In CRB-MAC (or CRB-RPL), any node in the forwarder set will erroneously forward the data packet if it does not hear the preamble or the data frame on the medium. The probability of this event (failure) is given by
\begin{equation}
\label{err_crb}
\mathcal{P}_f=P_{sw}^j \left[1-(1-p)^{m} \right]^{r_m} \cdot \left(1-(1-p)^d \right),
\end{equation}	 
where \(m\) and \(d\) denote the size (in bits) of micro-frame and data frame, respectively, \(r_m\) denotes the number of micro-frames in the preamble, and \(p\) denotes the bit error probability. Moreover, \(P_{sw}^j\) denotes the probability of switching transmission to the \(j^{th}\)  channel, which is given by (7) of \cite{R_CRB}. 

Unlike CORPL, there is no acknowledgement scheme between forwarding nodes in CRB-MAC (or CRB-RPL). Hence, the total path cost of sending a message from a sender node to the gateway/sink node depends on \cite{pot_OR}  (i) the cost of forwarding the message to the forwarder set, and (ii) the remaining path cost from the forwarder set to the gateway. Therefore, the total path cost to forward a message from node $i$ to gateway with forwarder set $\mathcal{F}_i$ is given by

\begin{equation}
\label{total_cost}
\begin{aligned}
\allowdisplaybreaks
Y_i=&\frac{1}{1-\prod_{j \in \mathcal{F}_i } (1-P_{ij})}  \\
&+\frac{ Y_1P_{i1}+\sum_{j=2}^{\mid \mathcal{F}_i \mid} Y_{j}P_{ij} \cdot \prod_{n=1}^{j-1} (1-P_{in}) }{1-\prod_{j \in \mathcal{F}_i } (1-P_{ij})},
\end{aligned}
\end{equation}
where it is assumed that nodes in $\mathcal{F}_i$ are sorted by their cost. Note that the denominator in the first and second terms of \eqref{total_cost} accounts for the probability that at least one node in $\mathcal{F}_i$ has received the message. Using \eqref{total_cost}, the coordination overhead for a node $i$ is given as follows.
\begin{equation}
\label{co}
\Delta_c^i=\sum_{b=1}^{\mid \mathcal{F}_i \mid} P_{ib} \cdot Y_{b} \cdot \prod_{r=1}^{b-1} (1-P_{ir}).
\end{equation}

It should be noted that the proposed approach is equally valid with a more generic definition (e.g., based on signal-to-noise ratio) of \(\mathcal{P}_f\).

\section{Comments on Performance Evaluation }

In Fig. 11 of \cite{R_smelly}, the authors evaluate  packet delivery ratio (PDR) performance of CRB-RPL against CORPL and RPL. The results show that CRB-RPL outperforms CORPL in terms of PDR. However, intuitively, there should not be any difference in performance of CORPL and CRB-RPL. This is because, for a fixed link success probability and a fixed number of nodes in the forwarder set, similar performance would be achieved whether the sender node prioritizes the forwarding nodes (CORPL approach), or the receiving nodes independently take part in the forwarding process (CRB-MAC approach). The opportunistic gain, from a fixed number of forwarding nodes, would be same in both cases.  




\section{Concluding Remarks}\label{sect_cr}
In this paper, we have identified the technical flaws associated with the protocol operation and analytical aspects of CRB-RPL. The authors' claim of CRB-RPL as an RPL-based routing protocol is not valid since the receiver-based forwarding approach does not retain the DAG structure of RPL. Moreover, CRB-RPL cannot guarantee protection to primary receivers due to the lack of DAG structure and the priority mechanism, and the dependence of CTQ metric on link success probability. Analytically, CRB-RPL does not specify some key parameters of interest and provides a technically incorrect approach to compute hop distance and coordination overhead. Besides highlighting the inaccuracies in analytical modeling, we have  provided an accurate approach, based on principles of opportunistic forwarding, to model the coordination overhead. Finally, the authors' claim of a new protocol is not justified as CRB-RPL adopts the operation of CRB-MAC.






\bibliographystyle{IEEEtran}

\bibliography{IEEEabrv,mybibfile}
%

\end{document}